\documentclass{ws-procs975x65}

\begin{document}

\title{LIEB--ROBINSON BOUNDS AND THE EXISTENCE\\ OF INFINITE SYSTEM DYNAMICS}

\author{BRUNO NACHTERGAELE}

\address{Department of Mathematics, University California, Davis\\
Davis, CA 95616, USA\\
$^*$E-mail: bxn@math.ucdavis.edu\\
\url{www.math.ucdavis.edu/~bxn}}

\begin{abstract}
We present a recent result on the existence of the dynamics in the
thermodynamic limit of a class of anharmonic quantum oscillator lattices,
which was obtained using Lieb--Robinson bounds.
\end{abstract}


\bodymatter

\section{Introduction}\label{sec:introduction}

In condensed matter physics, three common types of degrees of freedom
are often found together: atoms positioned in a lattice, spin magnetic moments, 
and itinerant electrons. Many phenomena primarily involve only one of these 
and it has therefore proved useful to study them with separate models.
Here, we are concerned with the spatial degrees of freedom of atoms in
a lattice. The fact that the atoms oscillate about their equilibrium positions is essential 
for many important phenomena in condensed matter physics. The question we address
is whether one can define the time evolution, consistent with the Schr\"odinger equation,
for an infinite assembly of quantum oscillators such as a crystal lattice. 

It is natural to start by considering the standard harmonic interaction, which should
be an accurate description when the displacements from the equilibrium positions
are small. It is well known, however, that some basic phenomena require that we 
consider anharmonic perturbations. Our approach applies to multi-body interactions
that fall off sufficiently fast but for clarity and space limitations we will
restrict ourselves to local and nearest neighbor anharmonicities here and  refer to 
our forthcoming paper for the general case \cite{nachtergaele2009b}. We will also
limit the discussion to oscillators organized in a translation-invariant fashion on a lattice.
To keep the notation simple we will work with one-dimensional oscillators at each lattice
site, but this is not essential.
With some straightforward modifications of the analysis our main result can be generalized
to non-translation invariant models defined on a graph satisfying a few natural assumptions.

\section{Harmonic lattice systems}

Let $\Lambda$ denote a finite subset of the $\nu$-dimensional hypercubic lattice $\mathbb{Z}^\nu$
and define Hamiltonians $H_\Lambda$ as self-adjoint operators on 
$\mathcal{H}_\Lambda=\bigotimes_{x\in\Lambda} L^2(\mathbb{R})$ by the following expression
\[
H_\Lambda=\sum_{x\in\Lambda} \frac{1}{2m}p_x^2 + \frac{m}{2}\omega^2 q_x^2 +V(q_x)
+\!\!\sum_{x,y\in\Lambda,\vert x-y\vert=1}\!\!\lambda(q_x-q_y)^2+\Phi(q_x-q_y)
\]
where $\omega,\lambda,\geq 0$, and $p_x$ and $q_x$ are the canonical momentum
and position operators for the oscillator at $x\in\Lambda$.
The Heisenberg dynamics, $\{\tau^\Lambda_t\}_{t\in\mathbb{R}}$,  is defined
by
\[
\tau^\Lambda_t(A)=e^{itH_\Lambda} A e^{-itH_\Lambda}, \quad A\in \mathcal{B}(\mathcal{H}_\Lambda).
\]
The algebra of observables is a tensor product 
\[
\mathcal{B}(\mathcal{H}_\Lambda)=\bigotimes_{x\in \Lambda} \mathcal{B}(\mathcal{H}_x)\equiv \mathcal{A}_\Lambda
\]
such that for  $X\subset \Lambda$, we have $\mathcal{A}_X\subset\mathcal{A}_\Lambda$, 
by identifying $A\in \mathcal{A}_X$ with $A\otimes\mathbb{I}_{\Lambda\setminus X}\in 
\mathcal{A}_\Lambda$.  Therefore, for $A\in\mathcal{A}_X$, $\tau^\Lambda_t(A)$ is 
well-defined for all $\Lambda$ containing $X$.

Our main concern is the existence of a limiting dynamics: do we have 
$\tau_t$ such that
\[
\tau_t^\Lambda(A)\longrightarrow \tau_t(A),\mbox{ as }\Lambda \uparrow \mathbb{Z}^\nu
\]
in a suitable sense? For an anharmonic lattice of classical oscillators a positive answer to
this question was obtained by Lanford, Lebowitz and Lieb \cite{lanford1977}.

It is well-known that in the case of quantum spin systems (i.e., bounded local Hamiltonians)
one can use Lieb-Robinson bounds to establish the existence of the thermodynamic
limit \cite{bratteli1997}. The essential observation is as follows.
Let $\Lambda_n$ be an  increasing exhausting sequence of finite volumes
with Hamiltonians of the form
\[
H_{\Lambda_n}=\sum_{X\subset \Lambda_n} \Phi(X)
\]
where $\Phi(X)=\Phi(X)^*\in\mathcal{A}_X$. Then, for $n>m$, one easily derives the bound:
\begin{equation}
\Vert \tau_t^{\Lambda_n}(A)- \tau_t^{\Lambda_m}(A)\Vert
\leq 
\sum_{x\in \Lambda_n\setminus\Lambda_m}\sum _{X\ni x}
\int_0^{\vert t\vert}\Vert [\Phi(X),\tau_t^{\Lambda_m}(A)]\Vert\, ds.
\label{cauchy}\end{equation}
Therefore, if we can show that the commutators in the integrand have sufficiently
small norms, it will follow that the finite-volume dynamics form a Cauchy sequence.
Estimates for such commutators were first derived by Lieb and Robinson \cite{lieb1972}.
For $A\in\mathcal{A}_X$ and $B\in\mathcal{A}_Y$, they proved a bound of the form
\[
\Vert [\tau_t (A), B]\Vert\leq C e^{-a(d(X,Y)-v\vert t\vert)}\, ,
\]
where $C, a$, and $v$ are positive constants and $d(X,Y)$ denotes the distance
between $X$ and $Y$. Estimates of this type are now commonly referred to as
Lieb-Robinson bounds \cite{nachtergaele2006a, hastings2006, nachtergaele2006b,
nachtergaele2007}. For anharmonic lattice systems Lieb-Robinson bounds were recently 
proved in Ref. \refcite{nachtergaele2009}, and this work builds on the results obtained there. 

For the oscillator lattices with $\lambda\neq 0$, the approach suggested by
Eq. \ref{cauchy} does not work due to the unboundedness of $(q_x-q_y)^2$.
But we note that  for the {\em harmonic} lattice, $\tau^\Lambda_t$ can be calculated explicitly
and can also be defined for the infinite lattice. One then sees, however, that 
$\tau_t^\Lambda(A)$ cannot converge in norm. The idea is  to consider the infinite anharmonic system as a limit of finite-volume perturbations of the infinite harmonic system.
We mention that, with a different approach, Amour, Levy-Bruhl, and Nourrigat have recently
obtained convergence results for certain models by introducing suitable
Sobolev-like norms for the observables \cite{amour2009}.

Rigorous perturbation theory of infinite systems is available if the unperturbed 
infinite system dynamics has a suitable continuity property \cite{bratteli1997}.
In our case this will be continuity for the weak operator topology, which follows from 
known properties and the exact solution of the harmonic lattice.

Up to a redefinition of the parameters, the harmonic lattice model Hamiltonian 
on a finite subset $\Lambda\subset \mathbb{Z}^\nu$, is
\[
H_\Lambda=\sum_{x\in\Lambda} p_x^2 + \omega^2 q_x^2 
+\sum_{\vert x-y\vert=1}\lambda(q_x-q_y)^2
\]
acting on $\mathcal{H}_\Lambda$.
The creation and annihilation operators are defined by
\[
a_x  =  \frac{1}{\sqrt{2}} \left( q_x \, + \, i p_x \right) \quad \mbox{and} \quad a^*_x =  
\frac{1}{\sqrt{2}} \left( q_x \, - \, i p_x \right),
\]
which satisfy the Canonical Commutation Relarions (CCR):
\[
[a_x, a_y] = [a_x^*, a_y^*] = 0 \quad \mbox{and} \quad [a_x, a_y^*] = \delta_{x,y} \quad \mbox{for all } x,y \in \Lambda_L \, 
\]
$H_\Lambda$ is a quadratic expression in $a_x, a^*_x$, which can
be diagonalized by a Bogoliubov transformation \cite{manuceau1968}. This is 
usually done in Fourier space resulting in:
\[
H_\Lambda=\sum_{k\in\Lambda^*}\gamma(k)(2b^*_k b^{}_k +1)
\]
where $\gamma(k)  =  (\omega^2 \, + \, 4 \lambda \sum_{j=1}^{\nu}  \sin^2(k_j/2))^{1/2}$.
The $b^*_k$ are creation operators for the eigenmodes of the system, which
also satisfy the CCR, and the ground state of $H_\Lambda$ is the corresponding vacuum. 
The time evolution is simply given by $\tau_t (b^*_k) = e^{-2i\gamma(k)t}b^*_k$.
It is useful for us to express the Bogoliubov transformations in real space:
for each $f: \Lambda_L \to \mathbb{C}$, set
\[
a(f) \, = \, \sum_{x \in \Lambda_L} \overline{f(x)} \, a_x,  \quad a^*(f) \, = \, \sum_{x \in \Lambda_L} f(x) \, a_x^*\, .
\]
and similarly for the $b$'s. Then, there are real-linear operators $U$ and $V$
such that
\[
a(f)=b(U^*f)-b^*(V^*f),\quad a^*(f)=b^*(U^*f)-b^*(V^*f).
\]
Using these relations, one finds a one-parameter
group, $T_t$, such that
\[
\tau_t(a(f))=\tau_t(b(U^*f)-b^*(V^*f))=a(T_t f),\quad \tau_t(a^*(f))=a^*(T_t f).
\]
Since the canonical operators ($p$'s, $q$'s , $a$'s, $a^*$'s,...) are all unbounded,
we will derive Lieb-Robinson bounds for the Weyl operators instead:
\[
W(f)  =  \mbox{exp} \left[ \frac{i}{\sqrt{2}} \left( a(f) \, + \, a^*(f) \right) \right] \, ,
\]
This can be done very explicitly by using the relations
\[
\tau_t(W(f))=W(T_t f), \mbox{ and } W(f)W(g)=e^{i\sigma(f,g)}W(f+g)
\]
where $\sigma(f,g)=\mbox{Im}\langle f,g\rangle$. Observe that
\begin{eqnarray*}
\left[ \tau_t(W(f)), W(g) \right]  & = & \left\{ W( T_t f) - W(g) W( T_tf) W(-g) \right\} W(g)  \\
& = & \left\{ 1 - e^{-2i \sigma( T_tf, g)} \right\} W( T_tf) W(g) \, .
\end{eqnarray*}
Since the Weyl operators are unitary, we therefore have
$\left\| \left[ \tau_t(W(f)), W(g) \right]  \right\| \leq 2\vert \sigma(T_t f,g)\vert$.
So, all we have to do is to estimate $\sigma(T_t f,g)$. 
$T_t f$ is explicitly given by 
$T_t f = f* \overline{h^{(L)}_{1,t}} \, + \, \overline{f} * h^{(L)}_{2,t}$, where
the functions $h_{1,t}$ and $h_{2,t}$ are given by
\begin{eqnarray*}
h^{(L)}_{1,t}(x) &=& \frac{i}{2} {\rm Im} [ \frac{1}{| \Lambda_L|} \sum_{k \in \Lambda_L^*}
 ( \gamma(k) + \gamma(k)^{-1} ) \, e^{ik \cdot x - 2i \gamma(k) t} ]  + {\rm Re} [  \frac{1}{| \Lambda_L|} \sum_{k \in \Lambda_L^*} e^{ik \cdot x - 2i \gamma(k) t} \, ]\\
h^{(L)}_{2,t}(x) &=& \frac{i}{2} {\rm Im} [ \frac{1}{| \Lambda_L|} \sum_{k \in \Lambda_L^*}
 (\gamma(k)  -  \gamma(k)^{-1} ) \, e^{ik \cdot x - 2i \gamma(k) t} \, ] .
\end{eqnarray*}
By analyzing these functions one then shows that for every $a>0$, there exist 
$c_a$ and $v_a$ such that
\[
\left| \sigma( T_t f, g ) \right| \leq c_a e^{ v_a |t|} \sum_{x, y} |f(x)| \, |g(y)| e^{-a d(x,y)}.
\]

\begin{theorem}[Ref. \refcite{nachtergaele2009}]\label{LR}
Let $\lambda,\omega \geq 0$. Then, for all  $f, g$ with $\,\mathrm{supp}\,  f \subset X$ and 
$\,\mathrm{supp}\,  g \subset Y$, 
\[
\Big\| \left[ \tau_t (W(f)) , W (g) \right] \Big\|
\leq  C\sum_{x,y}\vert f(x)\vert \, \vert g(y)\vert e^{- 2 (d(x,y) - v |t|) },
\]
with $v = 6\sqrt{\omega^2+4\nu\lambda}$.
\end{theorem}

\section{Anharmonic lattice systems}

In Ref. \refcite{nachtergaele2009} we proved Lieb-Robinson bounds for
finite systems, with estimates uniform in the volume. With the same 
approach one can also prove bounds for harmonic infinite systems
with an anharmonic perturbation in finite volume, i.e., a model with
formal Hamiltonian of the form:
\[
H_\Lambda=\sum_{x\in\mathbb{Z}^\nu} p_x^2 + \omega^2 q_x^2 \quad
+\sum_{x,y\in\mathbb{Z}^\nu, \vert x-y\vert=1}\lambda(q_x-q_y)^2+\sum_{X\subset\Lambda}\Phi(X).
\]
Typical $\Phi$ are $\Phi(\{x\})=V(q_x)$, $\Phi(\{x,y\})=W(q_x-q_y)$, etc.
but more general perturbations can be considered \cite{nachtergaele2009b}.

For the harmonic system, we can directly see how the exact expressions
extend to the infinite system. For $f:\mathbb{Z}^\nu\to\mathbb{C}$ in a suitable function space
($\ell^1(\mathbb{Z}^\nu)$, or $\ell^2(\mathbb{Z}^\nu)$), it is straightforward to define
\[
\tau_t(W(f))=W(T_t f)
\]
We get a Hilbert space representation by representing the Weyl operators
on the Fock space generated by the $b^*(f)$ operators acting on the 
vacuum. On this space we have well-defined Hamiltonian such that
\[
\tau_t(W(f))=e^{itH}W(f) e^{-itH}
\]
$\tau_t$ is the dynamics for the formal Hamiltonian
\[
H=\sum_{x\in\mathbb{Z}^\nu} p_x^2 + \omega^2 q_x^2 \quad
+\sum_{\vert x-y\vert=1}\lambda(q_x-q_y)^2
\]
and the Lieb-Robinson bounds continue to hold.

For simplicity, consider the perturbation of the form $P_\Lambda=\sum_{x\in\Lambda} V(q_x)$.
Then, the perturbed dynamics, formally corresponding to $H+P_\Lambda$, 
and can be defined mathematically by the Dyson series:
\[
\tau_t^{(\Lambda)}(W(f))= \tau_t (W(f))
+ \sum_{n=1}^\infty i^n\!\!\! \int_{0\leq t_1\leq t_2\cdots\leq t}
\!\!\!\!\!\!\!\!\!dt_1\cdots dt_n [\tau_{t_n}(P_\Lambda),[\cdots[\tau_{t_1}(P_\Lambda),\tau_t(W(f))]]]
\]
We have the following Lieb-Robinson bounds for $\tau_t^{\Lambda}$.
\begin{theorem}[Ref. \refcite{nachtergaele2009}]\label{thm:LRa}
Let $\lambda\geq 0,\omega > 0$, and $V$ such that
$\Vert k^2 \hat V (k)\Vert_1<\infty$.
Then, for all  $f, g\in\ell^1(\mathbb{Z}^\nu)$, we have
\[
\Big\| \left[ \tau^\Lambda_t (W(f)) , W (g) \right] \Big\|
\leq  C\sum_{x,y}\vert f(x)\vert \, \vert g(y)\vert e^{- 2 (d(x,y) - v |t|) }
\]
with
\[
v = 6\sqrt{\omega^2+4\nu\lambda} + c\Vert k^2\hat V (k)\Vert_1  .
\]
\end{theorem}

To show convergence, we estimate 
$\Vert \tau_t^{\Lambda_n}(W(f)) - \tau_t^{\Lambda_m}(W(f)) \Vert$, 
for $\Lambda_m\subset\Lambda_n$ by considering $\tau_t^{\Lambda_n}$ as a 
perturbation of $\tau_t^{\Lambda_m}$. This gives
\[
\tau_t^{\Lambda_n}(W(f)) = \tau_t^{\Lambda_m}(W(f)) + i \int_0^t \tau_s^{\Lambda_n} \left( \left[ P_{\Lambda_n \setminus \Lambda_m}, \tau_{t-s}^{\Lambda_m}(W(f)) \right] \right) \, ds \, ,
\]
Therefore
\[
\left\|  \tau_t^{\Lambda_n}(W(f)) - \tau_t^{\Lambda_m}(W(f))  \right\| \leq
 \sum_{z \in \Lambda_n \setminus \Lambda_m} \int_0^{|t|} \left\| \left[ V(q_z), \tau_{|t|-s}^{\Lambda_m}(W(f)) \right] \right\|  ds 
\]
By writing $V(q_x)=\int \hat{V}(p)W(p\delta_x)dp\, ,\quad W(p\delta_x)=e^{ipq_x}$,
we can then use Theorem \ref{thm:LRa} for $\tau_t^{\Lambda_m}$ to obtain the convergence.
The convergence is uniform on intervals $[-t_0,t_0]$. We also immediately get continuity in 
$t$ of the limiting dynamics by an $\epsilon/3$ argument.

\begin{theorem}[Ref. \refcite{nachtergaele2009b}]
Assume $\omega>0$, $\Vert k \hat V (k)\Vert_1< \infty, \Vert k^2\hat V (k)\Vert_1 <\infty$. 
For all $f\in \ell^1(\mathbb{Z}^\nu)$, and all $t\in\mathbb{R}$, the limit
\[
\lim_{\Lambda\uparrow \mathbb{Z}^\nu}\tau_t^{\Lambda}(W(f))=\tau^\infty_t(W(f))
\]
converges in the operator norm topology and the resulting
the dynamics is continuous in $t$ in the  weak operator topology.
\end{theorem}

\section*{Acknowledgments}
The work reported on in this paper was supported by the National Science
Foundation under Grant \# DMS-0605342 and FRG grants DMS-0757581, 
DMS-0757424,  and DMS-0757327.

\end{document}